\documentstyle[preprint,aps]{revtex}
\begin{document}
\draft
\title{Equation of state for polymer liquid crystals: theory and
experiment}
\author{H.H.Strey \thanks{permanent address: Department of Polymer
Science and Engineering, University of Massachusetts Amherst,
Amherst, MA 01003}, V.A. Parsegian and R. Podgornik
\thanks{on leave from Department of Physics,
Faculty of Mathematics and Physics, University of Ljubljana and
Department of Theoretical Physics, ``J.Stefan'' Institute, Ljubljana,
Slovenia}}
\address{National Institutes of Health, National Institute of Child 
Health and Human Developement, Laboratory of Physical and Structural 
Biology,
Bldg.12A/2041, Bethesda MD 20892-5626}
\date{\today}
\maketitle
\begin{abstract}
The first part of this paper develops a theory for the free
energy of lyotropic polymer nematic liquid
crystals. We use a continuum model with macroscopic elastic moduli
for a polymer nematic phase.
By evaluating the partition function, considering only harmonic
fluctuations, we derive an expression for the free energy of the
system. We find that the configurational entropic part of the free
energy enhances the effective repulsive interactions between the chains.
This configurational contribution goes as the fourth root of
the direct interactions. Enhancement originates from the
coupling between bending fluctuations and the compressibility of the
nematic array normal to the average director.
In the second part of the paper we use osmotic stress to measure
the equation of state for DNA liquid 
crystals in 0.1M to 1M NaCl solutions. These measurements
cover 5 orders of magnitude in DNA osmotic pressure.
At high osmotic pressures the equation of state, dominated by exponentially
decaying hydration repulsion, is independent of the
ionic strength. At lower pressures the equation of state is dominated by
fluctuation enhanced electrostatic double layer repulsion.
The measured equation of state for DNA fits well with our theory for 
all salt concentrations. We are
able to extract the strength of the direct electrostatic double layer
repulsion.  This is a new and alternative way of measuring
effective charge densities along semiflexible polyelectrolytes.

\end{abstract}
\section{Introduction}
In biology, there exists a class of bulk materials that provides
the cell's structural stability and ensures its integrity in multi-cell
environments. These materials are often made from biopolymers
with varying intrinsic stiffness and helicity such as: actin, fibrin, collagen,
polysaccharides, elastin, tubulin.  In most cases biology controls
elastic properties by controlled polymerization and depolymerization
of the monomers and by enzymatically controlled crosslinking of
polymer strands.  Often biopolymers, because of their intrinsic
stiffness, form liquid crystals at in vivo concentrations.  Because
these polymers are mechanically uniform, their phase behavior is
predominantly determined by the volume fraction of the polymer in its
solvent, rather than by temperature.  Such materials are called
lyotropic systems.  In contrast, polymeric liquid crystalline
materials used in industrial applications (like main-chain and
side-chain polymer liquid crystals) are most often thermotropic,
because of their flexible backbones.  Thermotropic means that
temperature determines their phase behavior.

Biopolymer materials are interesting for several reasons.  First of
all, there are many biomaterials made from or with biopolymers
exhibiting mechanical properties unreached by most conventional
synthetic materials, may the focus be strength, flexibility or a
combination of both.  Needless to say that little is understood about
their design, structure and how they work.  There is also the
increasingly appreciated potential of biopolymers for basic material
and condensed matter research \cite{livolantBP}.  In most synthetic
polymer systems it is extremely difficult to control properties of
individual polymers, like their degree of polymerization, crosslinks
and chemical uniformity.  By using modern molecular biological and
biochemical techniques it is possible to tailor biopolymers almost at
will, using nature's own efficient polymerization machinery.  So it
is, for example, possible to prepare monodisperse DNA fragment
solutions with lengths varying from a few $nm$ to several $\mu m$ using 
recombinant DNA methodology.  This is not yet possible with any other polymer.

DNA liquid crystals are
also of immediate interest, because they model
packing of DNA in confined spaces, as in cell nuclei and viral phage
heads (see e.g.  \cite{livolant,booy,strzeleckaN}).  Moreover, it is
believed that packing of DNA in chromatin plays an important role in
gene-regulation \cite{wolffe}.  By understanding how much energy is needed to
compact DNA we will gain insight into these processes as well.

One of the questions in condensed matter physics is how to
relate microscopic interactions to condensed state bulk properties,
like elastic and dielectric constants, heat capacity and packing
symmetries.  In an experiment we can go the other
way. The question is then: Can we infer microscopic interactions from
macroscopic behavior?

In this paper we explore the equation of state for DNA liquid crystals 
\cite{strey} using the osmotic stress method \cite{parsegian}.
It allows us to control all intensive
variables of the system, namely the chemical potentials of both water
and salt. According to the Gibbs-phase rule, this leaves us,
after equilibration against solutions of known activities, with a 
single phase.
The absense of phase coexistence is crucial for determining the
structure of the different phases using x-ray scattering.
At the same time the osmotic stress method provides us with information
about the free energy of the system, which can be obtained by 
integrating the equation of state (osmotic pressure $\Pi$ versus 
DNA density).

The ability to measure an equation of state and to map the phase
diagram for free energies made us realize that there was no
polymer liquid crystalline theory for free energies. This is not
surprising considering that typical liquid crystalline theories
are continuum elastic theories expanding deformations around
the symmetries of the system. Because these approaches are only good
for long wavelength physics, it is very hard to use them to predict free
energies.

Despite these difficulties, we will show in the first part of this 
paper that in our particular case with exponential interactions
between the polymer chains, the free energy can be derived starting
from a macroscopic theory. The microscopic interaction were 
incorporated into the continuum elastic moduli: the three Frank constants 
$K_{1}$, $K_{2}$, $K_{3}$ and the bulk compressibility
perpendicular to the chains $\cal B$.
By evaluating the partition function, considering only harmonic
fluctuations, we derive an expression for the free energy of the
system. To make the free energy finite we had to introduce a short
wavelength cutoff. We argue that the form of the free energy is
robust, meaning that the free energy behavior with respect to density
does not strongly depend on the particular choice of the cutoff.

We find that the fluctuation part of the free energy, going as the fourth root of
the direct interactions, enhances the repulsive interactions between the chains.
Enhancement originates from coupling between bending fluctuations and the
compressibility of the nematic array normal to the average director. The results should be
applicable to lyotropic polymer liquid crystals with fluid-like
positional order like nematics or hexatics. We think that it even can be applied
to chiral phases, like cholesterics and blue phases.  In typical
chiral phases the twist extends over several hundred intermolecular
spacings, so that in these phases local parallel packing can still be
used.

We compare theoretical predictions with measurements of equations of
state for DNA liquid crystals under different ionic condition over
almost five orders of magnitude in osmotic pressure \cite{strey}.  By
extracting the form of the fluctuation part of the measured free
energy and assuming exponentially decaying repulsive interactions
(screened electrostatic and hydration) between DNA molecules we have 
been
able to extract the strength of the direct electrostatic double layer
repulsion.  This extraction creates a new and alternative way of directly measuring
effective charge densities along semiflexibe polyelectrolytes.

\section{polymer nematic theory}
Let us first consider the elastic free energy of a nematic: a three
dimensional liquid with long-range orientational order and an average
director ${\bf n}$ along the z-axis.  Such phases are typically formed
by solutions of rod-like or disc-like objects.  There are three kinds
of deformations in quadratic order of ${\bf n}$ with symmetry
$C_{\infty h}$: splay, twist and bending.  The corresponding elastic
constants for these deformations are the Frank constants $K_{1}$ (splay),
$K_{2}$ (twist) and $K_{3}$ (bend).

\begin{equation}
\label{nematic}
{\cal F}_N = \textstyle{\frac{1}{2}} \int d^2{\bf r}_{\perp} dr_{z}
\left[ K_{1}\left(\nabla\cdot{\bf n}
\right)^{2} +  K_{2}\left({\bf
n}\cdot\left(\nabla\times{\bf n}
\right)\right)^{2} +  K_{3}\left({\bf
n}\times\left(\nabla\times{\bf n}
\right)\right)^{2}\right] \nonumber
\end{equation}

For small deviations of the director field
${\bf n({\bf r})}$ around its average orientation along the
$z$ axis ${\bf n({\bf r})} \approx \left( \delta n_{x}({\bf r}),
\delta n_{y}({\bf r}),
1\right)$, the free energy assumes the form

\begin{equation}
{\cal F}_N = \textstyle{\frac{1}{2}} \int d^2{\bf r}_{\perp} dz
\left[ K_{1}\left(\nabla_{\perp}\cdot{\delta\bf n}
\right)^{2} +  K_{2}\left(\nabla_{\perp}\times{\delta\bf n}
\right)^{2} +  K_{3}\left(\partial_{z}{\delta\bf n}\right)^{2}\right] \nonumber
\end{equation}

\bigskip

For polymer nematics we now have to consider that the director field ${\bf n({\bf r})}$ 
and the density of polymers in the (x,y)-plane $\rho=\rho_{0}+\delta\rho$ are
coupled \cite{degennesC,meyer}.  If the polymers were infinitely long and stiff
the coupling is given by the continuity equation:

\begin{equation}
	\partial_z \delta\rho + \rho_0 \nabla_{\perp} \cdot\delta{\bf n}=0
	\label{ }
\end{equation}

This constraint, however, is softened if the polymer has a finite
length $l$ or a finite persistence length ${\cal L}_{p}$.  On length
scales larger than $l$ or ${\cal L}_{p}$ the polymer can either fill
the voids with its own ends or fold back on itself \cite{semenov}. On
these length scales the polymer nematic can splay without density
change as illustrated in Fig.1.  Following \cite{doussal,kamien} this
can be expressed by introducing G, a measure how effectively the
constraint is enforced.  Density changes are expanded to second order
in density deviations $\delta\rho ({\bf r}_{\perp}, z) = \rho ({\bf
r}_{\perp}, z) - \rho_0$. $\cal B$ is the bulk modulus for 
compressions and dilations normal to the chains.
The total free energy can be written

\begin{equation}
{\cal F} = {\cal F}_{0}(\rho_{0}) + \textstyle{\frac{1}{2}} \int d^2{\bf
r}_{\perp} dz
\left[ {\cal B}\left(\frac{\delta\rho}{\rho_0}\right)^{2} + G\left(
\partial_z \delta\rho + \rho_0 \nabla_{\perp}\cdot \delta{\bf
n}\right)^2\right] + {\cal F}_{N} \nonumber
\end{equation}
where G is given by
\begin{equation}
	G = \frac{k_BT{\ell}}{2 \rho_0}
	\label{ }
\end{equation}
with $l$ never to exceed ${\cal L}_{p}$ \cite{kamienP}.
In the limit of finite polymer length G is also
finite and can be obtained from the observation that  $\partial_z
\delta\rho + \rho_0 \nabla_{\perp} \cdot\delta{\bf n}$ equals the
local difference between the number of chain heads and tails \cite{meyer}.
From here one derives that G is the concentration
susceptibility for an ideal mixture of heads and tails thus
\begin{equation}
	G = \frac{k_{B}T}{(\rho_{H}+ \rho_{T})},
	\label{ }
\end{equation}
where $\rho_{H}$ and $\rho_{T}$ are the average concentrations of
heads and tails, with $\rho_{H},\rho_{T} = \rho_{chain}$.  The chain
density on the other hand equals $\rho_{chain} =
\frac{\rho_{0}}{\ell}$, wherefrom $G = \frac{k_BT{\ell}}{2
\rho_0}$.
The corresponding structure factor can be written as \cite{kamien}

\begin{equation}
	{\cal S}(q_{\perp},q_{z}) = \mathopen< \vert
	\delta\rho(q_{\perp},q_{z})\vert^{2}\mathclose> =
	k_{B}T~\frac{\rho_{0}^{2}q_{\perp}^{2} + k_{B}T\frac{{\cal K}({\bf
	q})}{\cal G}}
	{{\cal B}q_{\perp}^{2} + k_{B}T\left( \frac{{\cal
	B}}{G\rho_{0}^{2}} + q_{z}^{2}\right){\cal K}({\bf
	q})},
	\label{ }
\end{equation}
where we defined
\begin{equation}
{\cal K}({\bf q}) = \frac{K_{1}q_{\perp}^{2} + K_{3}q_{z}^{2}}{k_{B}T}.
\end{equation}

For long-fragment DNA the limit ${\ell} \longrightarrow \infty$ is
appropriate, leading to the structure factor proposed by Selinger and 
Bruinsma \cite{selinger}
\begin{equation}
{\cal S}(q_{\perp},q_{z}) =
k_{B}T~\frac{\rho_{0}^{2}q_{\perp}^{2}}{K_{1}q_{\perp}^{2}q_{z}^{2} +
K_{3}q_{z}^{4}  +  {\cal B} q_{\perp}^{2}}.
\end{equation}

In order to calculate the contribution to the free energy due to
fluctuations in nematic order we have to sum over all the density
modes. From here on we will consider only fluctuations that are
coupled to density changes. Because we are interested in how the
free energy changes with density only those fluctuations will contribute.
Twist deformations do not couple to density
variations and can therefore be neclected.

\begin{equation}
{\cal F} = \textstyle{\frac{1}{2}}~k_{B}T\int\!\!\!\int
\frac{d^{2}q_{\perp}dq_{z}}{(2\pi)^{3}}~\ln
\left(K_{1}q_{\perp}^{2}q_{z}^{2} + K_{3}q_{z}^{4} + {\cal
B}q_{\perp}^{2}\right).
\label{crap1}
\end{equation}
To evaluate the integral we take the partial derivative with respect to
the compressibility $\cal B$
\begin{equation}
	\frac{\partial {\cal F}}{\partial{\cal B}} =
	\textstyle{\frac{1}{2}}~k_{B}T V~\int\!\!\!\int
	\frac{q_{\perp}dq_{\perp} dq_{z}}{{(2\pi)^{2}}}~
	\frac{q_{\perp}^{2}}{K_{1}q_{\perp}^{2}q_{z}^{2} + K_{3}q_{z}^{4} +
  	{\cal B}q_{\perp}^{2}}.
	\label{parb}
\end{equation}
The $q_{z}$ integral can be done straightforwardly and we remain with
\begin{equation}
	\frac{\partial {\cal F}}{\partial{\cal B}} =
	\textstyle{\frac{1}{2}}~k_{B}T \frac{V}{(2\pi)^{2}}\frac{\pi}{2}~
	\int\frac{q_{\perp}^{3}dq_{\perp}}{\sqrt{{\cal
	B}q_{\perp}^{2}}\sqrt{K_{1}q_{\perp}^{2} + 2\sqrt{{\cal
	B}K_{3}q_{\perp}^{2}}}}.
	\label{bull}
\end{equation}
This integral depends essentially on the upper cutoff for $q_{\perp}
= q_{\perp max}$ and we obtain
\begin{equation}
	\frac{\partial {\cal F}}{\partial{\cal B}} = k_{B}T \frac{V}{4\pi}
	\frac{{\cal B}K_{3}}{K_{1}^{2}\sqrt{{\cal B}K_{1}}}
	~F\left( \frac{q_{\perp max}}{2\sqrt{\frac{{\cal B}K_{3}}{K_{1}^{2}}}}
	\right),
	\label{ }
\end{equation}
where the function $F(x)$ has been defined as
 \begin{equation}
 	F(x) = \int_{0}^{x}\frac{u^{3/2}du}{\sqrt{1 + u}} =
 	\textstyle{\frac{1}{4}}\left(\sqrt{x}\sqrt{1 + x}(2x - 3) +
 	3~{\rm sinh^{-1}} \sqrt{x}\right) =
 	   \left\{
 	   \begin{array}{ll}
 	   \textstyle{\frac{2}{5}}x^{5/2} & \mbox{;$ x \ll 1$} \\
 	   \textstyle{\frac{1}{2}}x^{2} & \mbox{;$ x \gg 1$}
 	   \end{array}
 	   \right. .
 	\label{ }
 \end{equation}
From here we obtain the two limiting forms of the free energy as
\begin{eqnarray}
	{\cal F} \simeq  \frac{k_{B}T~V}{5\times
2^{3/2}\pi}~\sqrt[4]{\frac{{\cal
	B}}{K_{3}}}~q_{\perp max}^{5/2} + \ldots ~~~~~& ; &
q_{\perp
	max} \ll 2\sqrt{\frac{{\cal B}K_{3}}{K_{1}^{2}}}
	\label{quad} \\
	{\cal F} \simeq \frac{k_{B}T~V}{16\pi}~\sqrt{\frac{{\cal
	B}}{K_{1}}}~q_{\perp max}^{2} + \ldots ~~~~~& ; &
q_{\perp
	max} \gg 2\sqrt{\frac{{\cal B}K_{3}}{K_{1}^{2}}}.
	\label{doub}
\end{eqnarray}
Obviously the long-wavelength dependent physics is very complicated and depends
crucially on the values of typical polymer length and the ratios of
elastic constants. However it depends also on the $q_{\perp}$ cutoff.
We have either to eliminate the cutoff by including higher order terms
in the original Hamiltonian or to choose a meaningful
cutoff. Higher order terms will capture the short-wavelength physics
and remove the divergence.

In the following paragraph we want to show how this can be done for
the low density limit eq.~\ref{quad}.
We are aware that in a more thorough treatment one has to include all possible
higher order terms, but for right now we only include terms that
will make integral eq.~\ref{bull} convergent.
Letting ${\cal B}={\cal B}(q_{z}, q_{\perp}), K_{3}=K_{3}(q_{z},
q_{\perp}),K_{1}=K_{1}(q_{z}, q_{\perp})$ and taking into account the
symmetry of the elastic nematic free energy, one can easily show that
in order to make the integral in eq.\ref{bull} converge in the low 
density limit eq.~\ref{quad}, it is only necessary to expand  $\cal B$ as
\begin{equation}
{\cal B}(q_{z}, q_{\perp})={\cal B}_{0}\left( 1 + \zeta
^{2}q_{\perp}^{2} + \eta^{4}q_{\perp}^{4}\right).
\end{equation}
All the other expansion terms either simply lead to a renormalization of
$K_{1}$ (presumed to be small anyhow) or they do not
add up to the convergence of the free energy.

This form immediately leads to a density-density correlation function 
in the (x,y)-plane which
oscillates and decays exponentially \cite{zubarev}. Specifically the
correlation
function ${\cal S}({\bf \rho}, z)$ averaged over the length of the
polymers assumes the form
\begin{equation}
       \frac{1}{L}\int_{0}^{L}{\cal S}({\bf \rho}, z) =
       \frac{1}{4\pi {\cal B}_{0}
       \zeta^{2}}\left(\frac{\zeta}{\eta}\right)^{4}*~\Re e\left\{
       \frac{K_{0}(u_{1}\frac{\rho}{\zeta}) -
       K_{0}(u_{2}\frac{\rho}{\zeta})}{u_{1}^{2} - u_{2}^{2}}\right\},
       \label{here}
\end{equation}
where $u_{1,2}^{2}$ are the two complex zeros of $1 + u^{2} +
\left(\frac{\eta}{\zeta}\right)^{4} u^{4} = 0$, and $K_{0}(x)$ is the
modified Bessel-function.  The form
eq.\ref{here} of the transverse correlation function is a reasonable
ansatz for a liquid structure factor.  Evaluating the integral
(eq.~\ref{parb}) leads to the following expression
\begin{equation}
\label{higherorderfe}
	{\cal F} = {\cal F}_{0} +
\frac{k_{B}T~V}{16\sqrt{2}\pi}~\sqrt[4]{\frac{{\cal
	B}_{0}}{K_{3}}}~\zeta^{-5/2}
	\int_{0}^{\infty}\frac{u^{3/2}du}{\left(1 + u^{2}+{\left(
	\frac{\eta}{\zeta} \right) }^{4} u^{4}\right)^{3/4}}
\end{equation}
We essentially derived the same result as in eq.~\ref{quad}. The only
difference is that $q_{\perp max}$ has been replaced by $\zeta ^{-1}$.
$\zeta$ represents the correlation length in a liquid, which for a
typical liquid goes as the distance between next neighors.
Eq.~\ref{higherorderfe} suggests that instead of considering higher order terms we
can simply choose a cutoff proportional to the Brillouin
zone radius $q_{\perp max} \simeq \frac{\pi}{d}$, where $d$ is the
effective separation between the polymers in the nematic phase. This 
is a physically meaningful and appropriate cutoff because the
underlying macroscopic elastic model has, by definition, to
break down at wavelengths comparable to the distance between molecules.

Let us now see how our free energy scales for hard core repulsion
between chains.
First we take the limit of low density $q_{\perp max} \ll
2\sqrt{\frac{{\cal B}K_{3}}{K_{1}^{2}}}$. For an
elastic polymer $K_{3}$ has two additive terms, an intrinsic one and
an interaction contribution. The first one stems from
the elastic nature of the polymers themselves and has the form
$K_{3} \simeq k_{B}T {\cal L}_{p} \rho_{0}$, while the second term is the
generic
form valid for any steric interaction with a hard core $a$ {\sl i.e.}
$K_{3}\simeq k_{B}T/a$ \cite{degennes}.  We now use the Helfrich
self-consistent argument to evaluate the free energy due to elastic
fluctuations  in a hard core potential. It amounts to taking ${\cal B} =
V\frac{\partial^{2}{\cal F}}{\partial V^{2}}$, where $V$ is the volume of
the system which can be taken as $V = L \times d^{2}$, $L$ being the
length of the sample.
The crucial step is now to choose the cutoff which we set as $q_{\perp
max} \simeq \frac{\pi}{d}$. Inserting this into eq.\ref{quad} and taking into
account that the dominant behavior of $K_{3}$ will be
$K_{3} \simeq k_{B}T {\cal L}_{p}\rho_{0}$, we remain with
\begin{equation}
	{\cal F}(d) \sim L
	d^{2}\sqrt[4]{\frac{d^{4}}{k_{B}T}\frac{\partial^{2}{\cal
F}(d)}{\partial
	(d^{2})^{2}}}~d^{-5/2}.
	\label{ }
\end{equation}
The solution of this differential equation leads to the
scaling form for the fluctuation free energy {\sl i.e.} ${\cal F}(d)
\sim d^{-2/3}$.

In the opposite limit we need to estimate $K_{1}$
which is following de Gennes and Prost \cite{degennes} obtained as $K_{1} \sim
\frac{k_{B}T}{a}$, where $a$ is the hard core diameter of the polymers.
Thus in this case we obtain
 \begin{equation}
	{\cal F}(d) \sim L
	d^{2}\sqrt{\frac{d^{2} a^{2}}{k_{B}T}\frac{\partial^{2}{\cal
F}(d)}{\partial
	(d^{2})^{2}}}~d^{-2},
 	\label{ }
 \end{equation}
the solution of which leads to the scaling form ${\cal F}(d)
\sim d^{-2}$. By analyzing whether $q_{\perp
max}$ is smaller or larger then $2\sqrt{\frac{{\cal
B}K_{3}}{K_{1}^{2}}}$ it
is easy to ascertain that the first limiting law should be valid for
larger and the latter for smaller relative densities of the polymers.

Interestingly we just derived the same scaling laws as those
from
corresponding mean field models by Helfrich \cite{helfrich} and
De Gennes \cite{degennesP}.

\section{mean field models}

The mean field theory that corresponds to our problem
describes a single polymer
in a liquid crystalline matrix. In this view the polymer
fluctuates in the field of its neighbors. These mean field models
( one could also say ``Einstein'' cage models \cite{jain}) are strictly valid
only for a hexagonal LC phase where the
average position of a single polymer chain is well defined.
In nematic-like phases the mean-square displacements are
infinite and one has to choose a different language (eq.~\ref{nematic}).
Although mean field models are not the proper description
for our problem, they still illustrate the underlying physics. We
therefore review their results.

Let us imagine a
flexible (elastic) polymer in a hard tube of diameter $D$, oriented on
average along the axis of the tube, described with a Hamiltonian
of a persistent chain \cite{kratky}
\begin{equation}
	{\cal H} = \textstyle{{1}\over{2}}k_{B}T{\cal L}_{p} \int ds
	\left(\frac{d^{2}{\bf r}(s)}{ds^{2}}\right)^{2},
	\label{ela}
\end{equation}
where ${\cal L}_{p}$ is the persistence length of the polymer,
associated with its bending elastic constant $k_{c} = k_{B}T {\cal
L}_{p}$. For small deviations away from the average polymer position
along the z-direction the two-dimensional displacement vector ${\bf r}(s)$
can be
written as ${\bf r}(s)= (r_{x}(s), r_{y}(s))$. We assume that the diameter of
the tube is small enough (smaller than the persistence length ${\cal
L}_{p}$) so that between two consecutive hits with the walls of the
tube, the polymer propagates ballistically
\begin{equation}
	\mathopen< \left({\bf r}(s) - {\bf r}(s')\right)^{2} \mathclose>
	\simeq {\cal L}_{p}(s - s')^{2}.
	\label{inert}
\end{equation}
The longitudinal correlation length ${\cal L}_{\parallel}$ can be
obtained from evaluating the angle of the hit with the wall $\theta$
through statistics of a persistent chain {\sl i.e.}
  \begin{equation}
  	\theta^{2} \simeq \frac{D^{2}}{{\cal L}_{\parallel}^{2}} \sim
  	\frac{{\cal L}_{\parallel}}{{\cal L}_{p}},~{\rm wherefrom}~{\cal
  	L}_{\parallel} \sim {\cal L}_{p}^{1/3} D^{2/3}.
  	\label{ }
  \end{equation}
The free energy corresponding to the bumping between the polymer and
the wall on a lengthscale defined by ${\cal L}_{\parallel}$ can thus
be obtained as
\begin{equation}
	{\cal F} \sim \frac{k_{B}T}{{\cal L}_{\parallel}} \sim D^{-2/3}.
	\label{odijk}
\end{equation}
This result has been obtained previously by Odijk \cite{odijkLC} and
Helfrich \cite{helfrich} and represents a proper dimensional
generalization of the Helfrich fluctuational force to 1-D confined elastic
objects.

On much larger length scales where the polymer chain effectively
behaves as a free flight chain in 2-d, the effect of local nematic order can
be modeled through a simple nematic coupling term in the elastic
energy of the type $\left({\bf N}{\bf n}(s)\right)^2$ where $\bf N$ is
the average nematic director and ${\bf n}(s)$ is the local polymer
director, {\sl i.e.} ${\bf n}(s) = \dot{\bf r}(s)$.  If ${\bf N}$ is
directed again along the $z$ axis, the dominant part of the chain
Hamiltonian then assumes the form
\begin{equation}
	{\cal H} = \textstyle{{1}\over{2}}g \int ds
	\left(\frac{d{\bf r}(s)}{ds}\right)^{2},
	\label{nema}
\end{equation}
where $g$ gives the strength of the nematic coupling. This
Hamiltonian gives rise to a diffusive propagation of the polymer chain
described with
\begin{equation}
\mathopen< \left({\bf r}(s) - {\bf r}(s')\right)^{2} \mathclose> \simeq
g^{-1}(s - s'),
\label{ }
\end{equation}
if compared to the ballistic propagation of a persistent chain
Eq. \ref{inert}. For too large deviations the effective tube again
simulates the effects of steric interactions. The angle of the hit
with the wall of the tube is obtained as
\begin{equation}
\theta^{2}\simeq \frac{D^{2}}{{\cal L}_{\parallel}^{2}} \sim \frac{1}{g{\cal
L}_{\parallel}},~{\rm wherefrom}~{\cal L}_{\parallel} \sim
g~D^{2},
\label{ }
\end{equation}
where we have now used the statistics of a free flight chain. The
corresponding free energy is now obtained as
\begin{equation}
{\cal F} \sim \frac{k_{B}T}{{\cal L}_{\parallel}} \sim D^{-2}.
\label{ }
\end{equation}
Both results have already been derived for confined polymers 
\cite{degennesP}.
The point we wanted to make is that they are consistent with our
macroscopic theory if the cutoff is chosen appropriately, meaning
$q_{\perp max} \simeq d^{-1}$ while also equating the tube diameter
$D$ with the mean separation between the polymers $d$.

A similar approach can be used also in the case where confinement is
not described by a short-range hard-core interaction but is mediated
via a soft confinement potential of the form $V({\bf r}(z))$.  In this
case, assuming that the fluctuations of the polymer position from its
average are small, the Hamiltonian describing polymer elastic
fluctuations can be written in the form
\begin{eqnarray}
{\cal F} &=& \frac{1}{2} k_c \int_{0}^{L} dz
\left(\frac{d^2{\bf r}(z)}{dz^2}\right)^2 + \int_{0}^{L} dz \tilde V({\bf
r}(z)) \simeq \nonumber\\
&\simeq& \frac{1}{2} k_c \int_{0}^{L} dz
\left(\frac{d^2{\bf r}(z)}{dz^2}\right)^2 + \frac{1}{2} V'' \int_{0}^{L}
dz ~{\bf r}^2(z)
\end{eqnarray}
with
\begin{equation}
V'' = \nabla_{\perp}^{2}\tilde V({\bf r}(z)) = \left(
\frac{1}{r}\frac{d\tilde V({\bf r}(z))}{dr} +
\frac{d^2\tilde V({\bf r}(z))}{dr^2}\right) \vert_{r = D}
\end{equation}
where we have expanded the potential to second order in deviations
from a straight line.  One can define a length $\ell^*$ analogous to
the Odijk length by minimizing the energy with respect to the longitudinal
size $\ell$ of typical fluctuations \cite{jain}
\begin{equation}
{\cal F} \sim \frac{L}{\ell}\left( \frac{k_c~r^2}{\ell^3} + V'' r^2 \ell
\right) ~\longrightarrow~ \ell^{*4} \sim \left( \frac{k_c}{V''} \right).
\end{equation}
Thus we obtain for the confining energy
\begin{equation}
{\cal F}(D) \sim kT \frac{L}{\ell^*} \sim k_{B}T L {\left(
\frac{V''}{k_c} \right)}^{1/4}
\end{equation}
which apparently depends on the fourth root of the confining potential
stiffness $V''$. This result is completely consistent with the one
derived from a macroscopic theory eq.~\ref{quad},~\ref{doub} in the limit
$q_{\perp max} \ll 2\sqrt{\frac{{\cal B}K_{3}}
{K_{1}^{2}}}$, if we allow for the proper correspondence between the
confining potential stiffness $V''$ and the isothermal compressibility
modulus $\cal B$.

If on the other hand the polymer is presumed to be completely flexible with an
orientational part of the potential energy described with a nematic
coupling of the form eq.~\ref{nema}, the Hamiltonian of a single
confined chain will contain a softer elastic part proportional to
$\left(\frac{d{\bf r}(s)}{ds}\right)^{2}$.  In this case we have
\begin{equation}
{\cal F} = \frac{1}{2} g \int_{0}^{L} dz
\left(\frac{d{\bf r}(z)}{dz}\right)^2 + \frac{1}{2} V'' \int_{0}^{L}
dz ~{\bf r}^2(z)
\end{equation}
where the confining potential stiffness has been defined in the same
manner as before. Just as before we can again introduce the
appropriate Odijk length $\ell^*$ by minimizing the energy with respect to the
longitudinal size $\ell$ of typical fluctuations yielding \cite{rudi1}
\begin{equation}
{\cal F} \sim \frac{L}{\ell}\left( \frac{g~r^2}{\ell} + V'' r^2 \ell
\right) ~\longrightarrow~ \ell^{*2} \sim \left( \frac{g}{V''} \right).
\end{equation}
From here on by the same argument as before the confining energy scales as
\begin{equation}
{\cal F}(D) \sim kT \frac{L}{\ell^*} \sim k_{B}T L
{\left(\frac{V''}{g}\right)}^{1/2}
\end{equation}
therefore as the square root of the confining potential
stiffness. This result is now equivalent to the one derived from the
macroscopic theory eq.~\ref{quad},~\ref{doub} but this time in the limit
$q_{\perp max} \gg 2\sqrt{\frac{{\cal B}K_{3}}
{K_{1}^{2}}}$.

To summarize, the mean field theories surprisingly reproduce the
same scaling laws under similar conditions (high and low density 
regimes) as our more detailed calculation from part II. This indicates
that the fluctuation
part of the free energy is predominantely determined by
the entropy loss of confining the polymer chains. Still, in order
to model the free energy in more detail it is more 
appropriate to use our
macroscopic model, because it predicts the correct crossovers
between the different regimes.

The mean-field approach with nematic coupling constant $g$ as a free
parameter, inferred from experiments, was used for the first time in
\cite{rudi1} to argue that the effect of elastic fluctuations was to
renormalize the decay length of underlying exponential repulsion.
With the right choice of $g$, the calculated magnitude of the charge on
DNA agrees reasonably with the numbers derived below and in ref.
\cite{rudi2}.  Odijk \cite{odijkFE} also proposed what amounts to a
variation on the mean-field theme.  It is based on a variational
estimate for the Gaussian width of a single chain density distribution
function.  This theory in general does not lead to a straightforward
renormalization of the decay length of the underlying soft exponential
interactions between chains.  Also the phase boundaries calculated
from this theory coupled to Lindemann's criterion \cite{odijkFEL} fall
off the mark \cite{merchant}.

\section{Microscopic Interactions between DNA molecules}
In order to use our main result (eq.~\ref{quad},~\ref{doub}) for the
equation of state of nematic polymer liquid crystals we first have to
guess how the macroscopic elastic constants $K_{1} ,K_{2}, K_{3}$
and lateral compressibility $\cal B$ depend on the density ($\rho _{0}=1/A=\sqrt{3}/d^{2}$
for hexagonally packed chains). Our first guess is the ``one
constant approximation'' $K_{1}=K_{2}=K_{3}\approx U(d)/d$ \cite{degennes},
where $U(d)$ is the interaction energy and $d$ is the average distance
between the molecules. In our case this approximation has to be
modified, because the intrinsic bending stiffness $k_{c}$ of the DNA
molecules
contributes additionally to the bending Franks constant $K_{3}$.
\begin{eqnarray}
K_{1} & = & K_{2} \approx U(d)/d \nonumber \\
K_{3} & \approx & \rho_{0} k_{c} + U(d)/d \nonumber \\
B & \approx & V \frac{\partial ^{2}{{\cal F}_{0}(V)}}{\partial
V^{2}} = \frac{\sqrt{3}}{4 L} \left( \frac{\partial ^{2}{\cal F}_{0}}{\partial
^{2}d} - \frac{1}{d} \frac{\partial {\cal F}_{0}}{\partial d}
\right) \label{ks}
\end{eqnarray}

We now ask about the intermolecular interactions between two DNA
molecules.  There are two major contributions to the repulsion between
DNA molecules in monovalent salt solutions \cite{rudi2}: (1) screened
electrostatic repulsion from negative charges along the DNA backbone;
(2) hydration repulsion coming from partially ordered water close to
the DNA surface \cite{rau}. At the ionic strengths and polyelectrolyte
densities
considered in this work there appears to be no important contribution
to the attractive part of the total DNA-DNA interaction: van der
Waals forces are negligible \cite{rau} and counterion-correlation
forces \cite{oosawa,barrat} are screened \cite{rudiCF}.

The mean-field electrostatic interaction is best described by the
Poisson - Boltzmann theory resulting in an interaction potential
between two parallel charged rods \cite{brenner} of the form
\begin{equation}
\label{lcdens}
U(d)/L=\frac{\xi ^{2}}{2 \pi \epsilon \epsilon _{0}} K_{0}(d / \lambda_{D})
\label{tworods}
\end{equation}
where $\lambda_{D}$ is the Debye screening length.  For ionic
strengths $I$ from monovalent salts, $\lambda_{D}= 3.08\AA /
\sqrt{I[M]}$.  Eq.~\ref{tworods} refers to two infinitely thin line
charges with a charge density per length $\xi$.  This line charge
density is related to the actual surface charge density $\sigma$ on a
cylinder with radius $a$ as follows
\begin{equation}
\label{scdens}
\xi = 2 \pi \sigma \lambda_{D} / K_{1}(a/\lambda_{D})
\end{equation}
For large $d/\lambda_{D}$ the Bessel function $K_{0}$ can be
approximated by
\begin{equation}
K_{0}(d/\lambda_{D}) \approx \sqrt{\frac{\pi}{2}}
\frac{e^{-d/\lambda_{D}}}{\sqrt{d/\lambda_{D}}}
\end{equation}

The hydration repulsion between solvated molecules in water can be
described by the same formalism as for screened electrostatic repulsion
(see \cite{leikin}). The interaction energy therefore goes as
$K_{0}(d/\lambda _{H})$ with $\lambda _{H}\approx 3\AA$ \cite{rudi2}.

Knowing about the intermolecular interaction we can now revisit our main
result eq.~\ref{quad},\ref{doub}. From eq.~\ref{ks} it is clear
that $K_{1}$, $K_{2}$ and the bulk compressibility modulus 
perpendicular to the 
chains $\cal B$ essentially decay exponentially
with $d$ whereas $K_{3}$ after an initial exponential decay will
at low densities be dominated
by the intrinsic bending stiffness $k_{c}$ of the polymers.
This fact has instructive consequences.
For lower densities, where we would expect fluctuations to be more
prominent, only the limiting form valid for $q_{\perp max} \ll
2\sqrt{\frac{{\cal B}K_{3}}{K_{1}^{2}}}$ remains.
In this limit, using exponentially decaying interactions for $\cal B$,
the fluctuation part of the free energy goes essentially as the fourth root of
the direct interaction, since $K_{3}=\rho _{0} k_{c}$.

Summarizing, the bare interaction can be described in the following way
(for simplicity we use the free energy per length ${\cal G}={\cal F}/L$).
\begin{equation}
\label{barefe}
{\cal G}_{0}(d) = a \sqrt{\frac{\pi}{2}}
\frac{e^{-d/\lambda_{H}}}{\sqrt{d/\lambda_{H}}} + b \sqrt{\frac{\pi}{2}}
\frac{e^{-d/\lambda_{D}}}{\sqrt{d/\lambda_{D}}}
\end{equation}
where a and b are the amplitudes of the hydration and screened
electrostatic repulsion.
The total free energy ${\cal G}$, using the low density limit
eq.~\ref{quad} using $q_{\perp max}=\pi /d$, is then
\begin{equation}
\label{totalfe}
{\cal G}(d) = {\cal G}_{0}(d) + k_{B}T \frac{ {(\pi)}^{5/2}
}{20 \pi \sqrt{3}} k_{c}^{-1/4} \sqrt[4]{ \frac{\partial ^{2}{\cal
G}_{0}}{\partial
^{2}d} - \frac{1}{d} \frac{\partial {\cal G}_{0}}{\partial d} }
\end{equation}

\section{Materials and methods}
Sample preparation and determination of interaxial separations between
DNA molecules by either x-ray or direct density measurements was described
in detail previously \cite{strey}.

At very low osmotic pressures (1/100 atm) the elasticity of the
dialysis bags could contribute to osmotic pressure that acts on the
sample. For this reason we performed
experiments in which we dialysed low concentration ($0.1 wt\%-1wt\%$)
Dextran solutions against each other, to make sure that there were
no residual osmotic pressures resulting from partially inflated dialysis tubes.
After equilibration the dextran concentrations inside and outside the
dialysis bags agreed within $1\%$ of the bathing concentration down to $0.1
wt\%$
Dextran.

To compare our experimental results with our theory we
expressed all data in terms of the interaxial spacing between two
DNA molecules. Assuming hexagonal packing in all density regimes the
relation between density and the interaxial spacing $d$ is
$\rho=(610/d[\AA])^{2})[mg/ml]$.

The relation between the osmotic pressure $\Pi$ and $\cal{G}$ is then 
\cite{rau}:
\begin{equation}
	\frac{\partial \cal{G}}{\partial d}=\sqrt{3}\Pi d
	\label{*}
\end{equation}

The nonlinear fits were done using the Levenberg-Marquardt method,
implemented in the data analysis software Igor 3.03 (WaveMetrics, OR).
The fit function used was (using ${\cal G}_{0}$ from eq.~\ref{barefe}):
\begin{equation}
\label{fitfunc}
\frac{\partial \cal G}{\partial d} = \frac{\partial {\cal
G}_{0}}{\partial d} + c k_{B}T k_{c}^{-1/4} \frac{\partial}{\partial
d}\sqrt[4]{ \frac{\partial ^{2}{\cal
G}_{0}}{\partial
^{2}d} - \frac{1}{d} \frac{\partial {\cal G}_{0}}{\partial d} }
\end{equation}
where a and b are the bare amplitudes of the hydration and the
screened electrostatic repulsion. Since the prefactor of the
fluctuation part of the free energy depends on the cutoff wave-vector
we chose to fit it by the dimensionless constant c. The actual fit
was performed using $log_{10}(\partial{\cal G}/\partial d)$ versus
interaxial spacing $d$. We did that to achieve an equal weight of all data
points
over the whole osmotic pressure regime ( the osmotic pressure goes
roughly exponentially in PEG-concentration ).

\section{experimental results}

Fig.  2 shows the equation of state ($\Pi - d$) for long DNA molecules
at different NaCl concentrations (10mM - 2M).  The figure presents
a compendium of data (from \cite{rau}, \cite{rudi2}, \cite{rudi3}
and heretofore unpublished data) obtained up to date on the upper portion of
the DNA phase diagram. At high osmotic
pressures ($\Pi $) all ionic strengths merge into the same curve: an
exponential decay with a decay length of about $3\AA$.  We attribute
this behavior to structural forces in water (hydration forces)
commonly observed between hydrated surfaces in water \cite{leikin}.
At lower osmotic pressures the curves start to deviate from each other
reflecting the influence of screened electrostatic repulsion.
Interestingly for ionic strengths $\geq 1M$ the curves are
independent of ionic strength over the whole osmotic pressure-regime.
This indicates that for $I>1M$ the electrostatic contributions are
sufficiently screened so that the equation of state is dominated by
hydration repulsion alone.

Fig. 3 shows the measured $\partial{\cal G}/\partial d = \sqrt{3} \Pi d$ at
$1M$ NaCl. Since from $1M$ on the electrostatic contribution is
negligible we can use the data to determine the decay length of the
hydration repulsion. We fitted the data according to eq.~\ref{fitfunc} using
only one exponentially decaying direct interaction. As fit parameters
we used the amplitude of the hydration repulsion $a$, the hydration
decay length $\lambda_{H}$ and the prefactor $c$ of the fluctuation
part of the free energy. The fit is shown as a solid line and
describes the data very well. The resulting hydration decay length was
$\lambda_{H}=(2.9 \pm 0.2)\AA$. The amplitudes and the prefactor c are
summarized in table 1.

Fig. 4 and Fig. 5 show measured $\partial{\cal G}/\partial d$ for 0.5M
and 0.1M NaCl. The solid line shown is the fit to eq.~\ref{fitfunc}
using a,b and c as fit parameter. The hydration decay length was set
to $\lambda_{H}=2.88 \AA$ and the Debye screening length was set to
$\lambda_{D}=3.08\AA/\sqrt{I[M]}$. The results are shown in table 1.

Fig. 4 indicates the phase boundaries between the various liquid
crystalline phase of DNA in detail. For all other salt concentrations
we only indicated the isotropic to anisotropic transition.
At very high osmotic pressures (not shown in Fig.4) there exists a
crystalline hexagonal phase of DNA (see \cite{langridge}). It melts
into a line hexatic phase (regime a): a three dimensional liquid with
long-range
bond-orientational order perpendicular to the axis of the molecules
\cite{rudi4}. As far as we can see, there is no indication for a
hexagonal liquid crystalline phase in between the crystalline and
the line hexatic phase.
Between the line hexatic and the chiral phases of DNA the x-ray
structure factor shows two peaks (regime b): a sharper peak
(continuing regime a) at smaller interaxial spacings and a more
diffuse one at wider spacings (continued by regime c).
Since by using the osmotic stress method we hold all intensive variables
($p, T, \mu$) fixed the mesaured S(q) should originate from a
single phase (Gibb's phase rule). At this point it is not clear
whether the structure in regime b corresponds to a new phase
in between a line hexatic phase (nonchiral) and a cholesteric phase.

The chiral phases start with a cholesteric phase as observed by
electron microscopy and polarization microscopy \cite{leforestierCP,rudi5}.
At even lower concentrations there is some indication for additional
chiral phases (precholesteric phase \cite{livolantPC} and blue
phases \cite{leforestier}). Finally the anisotropic liquid crystalline
phase melts into an isotropic phase \cite{merchant}(regime d).

\section{discussion}

DNA is highly charged ( two negative charges per base-pair or $3.4\AA$
of its length). In an electrolyte solution the DNA¹s net negative charge
creates an accumulation of counterions close to its surface that
screen part of the bare charge and lead to an ³effective² charge density
that is felt at long distances between chains. Theoretically this effect
can be captured by nonlinear Poisson-Boltzmann theory 
\cite{stigternlpb}.

Even though our nematic polymer liquid crystalline theory was based
on a rather simplistic model it describes all the data fairly well
with reasonable values for the fitted parameters. In table 1
we have summarized the corresponding line-charge densities and
surface-charge-densities that were calculated according to
eq.~\ref{lcdens} and eq.~\ref{scdens}.  At 0.5M and 0.1M the fits give the
same surface-charge density $\sigma =0.07 ~C/m^{2}$. This value corresponds
to about $50\%$ of the bare charge of DNA ($0.15~C/m^{2}$).

Our result of $50\%$ effective
charge agrees very well with the analysis of electrophoretic
measurements \cite{ross} by Schellman and Stigter \cite{schellman}
that resulted in about $60\%$ effective charge density for Na-DNA.  A
recent study using steady-state-electrophoresis reported $10\%$ for
Na-DNA \cite{laue}.  This discrepancy exists not because of experimental
uncertaintly, but because of different theoretical treatments of the
measured mobilities (for a recent discussion see \cite{hoagland}).
Part of the difference might come from the complicated details of frictional forces
acting on a rough, charged polyelectrolyte.

One has to bear in mind that for
the $0.5M$ data the uncertainty of the fitted amplitude is almost 100\%.  
This uncertainty is not
very surprising considering the close proximity of the two decay
lengths $\lambda_{H}=2.9\AA$ and $\lambda_{D}=4.36\AA$.  As soon as the
decay lengths separate from each other, as in the case for 0.1M, the
statistical uncertainty drops to 10\%.

We also fitted the prefactor c. If in the theory (eq.~\ref{totalfe})
one chooses the cutoff wave-vector
to be at the Brillouin-zone radius $q_{\perp max}=\pi /d$, c evaluates to $0.16$.
The fitted prefactors c range from $1.3$ to $0.8$. This is about 5-8
times larger than the theory predicted. On the other hand, considering
the simple underlying model, the value is not too far off. Choosing a
cutoff at twice the Brillouin-zone radius, for example, would give a
value right on the fitted one. In our view, the prefactor depends
on the fine details of the in-plane structure-factor $S(q_{\perp})$ (see 
eq.~\ref{higherorderfe}). In the case of exponential direct interactions 
between the chains any
algebraic dependence of the cutoff with respect to the density
will be dominated by the fourth root of the
direct interaction (eq.~\ref{totalfe}). The fact that for different
ionic strength the prefactors have similar values ( all around 1 )
strengthens our argument.

The fits even seem to prove explicitly the presence of hydration
repulsion. If one
tries to calculate the charge densities for 1M, using a screening
length of $\lambda_{D}=3.08\AA$ and ignoring any contribution of
hydration repulsion, the surface charge density
results in $\sigma =0.19~C/m^{2}$, 25\% more charge than the total
phosphorus
charges on DNA. Since it is generally believed that more than half
of the bare charge on DNA is screened, a picture with pure electrostatic
double layer repulsion is hard to envision. Recently
Lyubartsev and Nordenski\"{o}ld \cite{mcdna} published Monte-Carlo
simulation addressing the DNA
case,
comparing their results with osmotic stress measurements from our
lab. They showed that electrostatic repulsion strongly increased
beyond simple linearized PB theory, when two charged cylinders
approach each other closely. 
However this simulation used a solid cylinder-model of DNA and hard
sphere $d=4\AA$ ions. The measurements are in a regime where most
of the aqueous volume is inside the DNA grooves rather than outside
any DNA cylinder. Better models have to include the possibility of 
ions entering the groove space.
Although their results seem to fit
the data for 0.5M quite well, there is not as good success at other
salt concentrations. We take the force data in high salt concentration
to be strong evidence for hydration repulsion.
Even non-charged polymers, like the polysaccharide schizophyllan show
an exponentially decaying repulsion with a decay length of about
$\lambda_{H}=3.4\AA$ \cite{raus}. Hydration repulsion
is a general feature of water soluble molecules at separations
$\leq 1nm$ \cite{leikin} that can not be simply denied \cite{jacob}.

Osmotic stress measurements can be used to determine
more directly effective charge densities of
semiflexible polyelectrolytes. We can
therefore test and compare theories that predict effective charge
densities, like non-linear Poisson-Boltzmann (PB) or Manning theory
\cite{manning}.
Our results indicate that for Na the effective charge is twice as
large as predicted by both theories.  Previous studies in our lab
\cite{rudi2}
observed significant differences in charge densities using different
counterions, such as Li,Na,K,Cs,Tri-methyl-ammonium.  These results
merit further analysis.

The form of the fluctuation part of the free energy (eq.~\ref{totalfe}) suggest
that fluctuation enhanced repulsion may be important for many
lyotropic polymer liquid crystalline systems.
The prefactor $c$ of the fluctuation part only depends weakly on the
bending constant of the polymer $c \propto k_{c}^{-1/4}$. The only
condition for enhanced repulsion is that the polymers are longer than
their persistence length, and that they remain in a nematic-like phase.

Another appealing conclusion from our work is the emerging
connection between multilamellar lipid and columnar polyelectrolyte
(DNA) arrays.  Both are governed by the same type of colloidal forces
\cite{parsegianR}, except that for polyelectrolyte arrays the
attractive forces at relevant ionic strengths and polymer
concentrations are usually negligible, and show the same
conformational flexibility describable by an elastic term in the
conformational Hamiltonian (Helfrich \cite{helfrichH} in the case of lipid
multilayers and Kratky-Porod \cite{kratky} in the case of
semiflexible chains).  The interplay between fluctuations and effective
interaction
in multilamellar arrays has been a topic for quite awhile
\cite{lipowsky} and is at present reasonably well understood.

Setting aside the attractive part in the bare interaction potential
and the different dimensionality of the fundamental interacting
objects (2-D in the case of lipid multilayers {\sl vs.} 1-D in the
case of polyelectrolytes), multilamellar lipid and columnar
polyelectrolyte arrays share exactly the same Hamiltonian.  This leads
to the same type of fluctuational renormalization of interactions
force that can be expressed by the following two forms of free energy
\begin{equation}
{\cal F} \simeq  \frac{k_{B}T~V}{5\times
2^{3/2}\pi}~\sqrt[4]{\frac{{\cal
	B}}{K_{3}}}~q_{\perp max}^{5/2} + \ldots ~~~~~vs.~~~~~
{\cal F} \simeq  \frac{k_{B}T~V}{16\pi}~\sqrt{\frac{{\cal
	B}}{K_{3}}}~q_{\perp max}^{2} + \ldots,
\label{}
\end{equation}
The first one pertains to the fluctuations in polyelectrolyte
arrays, {\sl c.f.} eq. \ref{doub}, and the second one to the
fluctuations in multilamellar systems, {\sl e.g.} \cite{helfrich2}. Both 
these results are
written in the limit of small density of fluctuating objects. The
apparent formal differences between the two expressions are due {\sl
solely} to the dimensional difference between the fluctuating
objects, {\sl i.e.} 1-D as opposed to 2-D.

The main difference between multilamellar and columnar arrays
interacting through exponential repulsive forces would thus be the
two-fold {\sl vs.} four-fold renormalisation of the decay length.

There have recently (see \cite{schmitz} for an overview), and maybe not so
recently \cite{sogami}, appeared quite a few speculations on the possible
attractive component to the polyelectrolyte interaction forces.  At
polyelectrolye densities and ionic strengths described in this work,
there is certainly {\bf no} evidence to presume there are any.  We can
not, however, exclude the possibility that low ionic strengths and low
polyelectrolyte densities, thus promoting pronounced unscreened
counterion fluctuations \cite{robjin,rudiCF}, conspire to bring
forth non-negligible attractive forces between DNA molecules.  Should
this turn out to be the case, columnar polyelectrolyte arrays would
become even more similar to multilamellar lipid arrays. DNA and
cell membranes are among the principal organizational structures in biology.
That they share such a pronounced amount of
common physics certainly add up to a rather pleasing
intellectual development.

\vskip 2 cm

{\bf Acknowledgement:} We want to thank Don Rau, David Nelson,
Robijn Bruinsma and Per Lyngs Hansen for fruitful discussions and Gary Melvin and Leepo
Yu for their support enabling the x-ray experiments. Special thanks
to Randy Kamien for many
delightful discussions and valuable comments. We would like to thank the Aspen
Institute for  Physics for the hospitality during the Summer 1996 workshop on
``Topological Defects in Condensed Matter Physics''.

\vfill
\eject

%
%

\begin{figure}
\caption{Illustration of how splay couples to the density of
different nematic liquid crystals: a) stiff, long rods; b) stiff, short 
rods; c) semiflexible, long polymers. Long, stiff rods (a) show strong density
changes when splayed. In the case of short and stiff rods (b) the voids created by
splaying the material are filled by other short rods.  For long, semiflexible
polymers (c) the voids are filled by polymers folding back on itself.}
\label{fig1}
\end{figure}
\begin{figure}
\caption{Equation of state for DNA liquid crystals at ionic strength
from 150mM to 2M NaCl. We plotted $log \Pi$ versus interaxial
spacing d. The interaxial spacing d was measured by x-ray scattering.}
\label{fig2}
\end{figure}
\begin{figure}
\caption{Measured $\partial{\cal G}/\partial d$ for DNA liquid crystals
at ionic strength of 1M over 5 orders of magnitude in osmotic pressure.
The solid curve represents a fit of the datapoints in the anisotropic
regime
to a nematic liquid crystalline theory
(see text) assuming exponential repulsion (hydration and
screened electrostatics). The broken line represents the bare
interaction without fluctuation enhanced repulsion.
Interaxial spacings above the -.- line are measured by x-ray scattering;
below this line the spacing d are derived from measured DNA densities
where molecules are expected to be hexagonally packed.}
\label{fig3}
\end{figure}
\begin{figure}
\caption{
Measured $\partial{\cal G}/\partial d$ for DNA liquid crystals
at ionic strength of 0.5M over 5 orders of magnitude in osmotic pressure.
Four structural regimes (a-d) could be distinguished. See Fig.3 for
annotations.
}
\label{fig4}
\end{figure}

\begin{figure}
\caption{Measured $\partial{\cal G}/\partial d$ for DNA liquid crystals
at ionic strength of 0.1M. See Fig.3 for annotations.}
\label{fig5}
\end{figure}

%
%

\vfill
\eject
\begin{table}
\caption{summary of all fitted parameters for 1M, 0.5M and 0.1M NaCl,
as well as the corresponding line charge density $\xi$, surface charge
density $\sigma$ and the fraction of unscreened phosphorus charges $E$}
\label{table1}
\begin{tabular}{|l|c|c|c|c|c|c|c|c|}
$I [M]$ & $a [J/m]$ & $\lambda_{H} [\AA]$ & $b [J/m]$ & $\lambda_{D} [\AA] $
& $c$ & $\xi [C/m]$ & $\sigma [C/m^{2}]$ & $E$ \\ \hline
1 & $(1.7\pm 0.9) 10^{-7}$ & $(2.9 \pm 0.2)$ & & 3.08 & $(1.2\pm 0.1)$ & &
&\\ \hline
0.5 & $(1.4\pm 0.5) 10^{-7}$ & $ 2.88 $ & $(3 \pm 2.6) 10^{-9}$ &
$4.36$ & $(1.3\pm 0.2)$ & $2.1\cdot 10^{-9}$ & 0.07 & $0.49$ \\ \hline
0.1 & $(1.1\pm 0.3) 10^{-7}$ & $2.88$ & $(4.1\pm 0.3)
10^{-10}$ & 9.74 & $(0.8\pm 0.06)$ & $ 7.8\cdot 10^{-10}$ & $0.07$ & 0.48\\
\end{tabular}
\end{table}

\end{document}